\documentclass[usenatbib]{mn2e}
\usepackage{graphicx}
  
\begin{document}
 
\title{Study of star formation in NGC 1084}
\author[S. Ramya et.al.]{S. Ramya 
\thanks{E-mail : ramya@iiap.res.in; dks@iiap.res.in; tpp@iiap.res.in},
D. K. Sahu, T. P. Prabhu\\
 Indian Institute of Astrophysics, Koramangala, Bangalore 560 034, India.}
 
 
\date{Received / Accepted}

\pagerange{\pageref{firstpage}--\pageref{lastpage}} \pubyear{2006}

\maketitle
 
\begin{abstract}
We present  $U$$B$$V$$R$$I$ broad band, H$\alpha$ narrow band photometry of the star forming complexes in the infra-red bright galaxy NGC 1084. Results of medium resolution spectroscopy of some of the brighter complexes are also discussed.  Spectroscopic data is used to better estimate the internal reddening within the galaxy which is found to be highly variable and to calculate metallicity which is close to the solar value. Diagnostic diagram identifies the shocked regions within this galaxy. The  narrow band H$\alpha$ flux  and its equivalent width are used to determine the star formation rates of the complexes and the distribution of ages. Star formation rates for a few of the complexes are found to be as high as 0.5 $M_{\odot}${yr}$^{-1}$. The star forming complexes lie in the age range 3 Myr to 6.5 Myr. \  $U-B$ vs $V-I$ colour-colour mixed population model created using the Starburst99 model colours is used to estimate the ages of the stellar populations present within these regions. Using this technique, it is found that the star formation in NGC 1084 has taken place in a series of short bursts over the last 40 Myr or so. It is proposed that the likely trigger for enhanced star formation is merger with a gas rich dwarf galaxy.

\end{abstract}

\section{Introduction}
\label{intro}
Although current theoretical models of star formation histories of galaxies are largely supported by powerful multiwavelength studies, further observational data still remains crucial for understanding chemical evolution, star formation history and the triggering mechanisms for star formation in galaxies. The effects of internal disk structure such as bars and spiral arms play a vital role, but it is the external environmental influences such as tidal interactions and mergers which have much stronger effects on star formation (review by \citealt{moorwood} and references therein). Numerous observations have established this link between strong starbursts and merger of galaxies. Several studies of such spectacular systems have been done, few of them being the Antennae (\citealt{bastian}, \citealt{hibbard}, \citealt{kassin}), NGC 6090 \citep{sugai}, NGC 7252 (\citealt{miller}, \citealt{ss}), NGC 5194 \citep{calzetti05}.

Galaxies evolve  not only dynamically but also chemically. Emission line spectra provide estimates of abundance of light elements which are useful towards a study of chemical evolution of galaxies. Dust distribution is highly chaotic in galaxies and may be uniform and clumpy in the foreground, as well as mixed with the emitting region \citep{calzetti94}. An accurate estimate of reddening is necessary for an accurate determination of the emission line ratios which provide an insight into the chemical evolution and help in estimating star formation rates.

In this paper we try to investigate the recent star formation and the star formation history in the spiral galaxy NGC 1084 using narrow band H$\alpha$, broad band $U$$B$$V$$R$$I$ and spectroscopy after obtaining reasonable estimates of the variable internal reddening within the galaxy.

NGC 1084 is classified as SA(s)c in RC3 catalogue \citep{devau}. Its classification has recently been revised by \cite{eskrid} based on the  Ohio State University (OSU) Bright Spiral Galaxy Survey, and is assigned the class of Sbc and Sb based on the  $B$ band and $H$ band images, respectively.   The $H$ band image of this galaxy reveals a bright nuclear point source, elongated bulge along the major axis, multi arm pattern, without well defined arms, and a number of bright knots.  This galaxy is known to have strong nuclear  and patchy circumnuclear H$\alpha$ emission \citep{knapen}. It possesses kinematically distinct regions, \cite{afanasiev} first detected  non-circular phenomena  in the central region (R $< 5''$) and in the periphery of the galaxy at  around $40''-\ 50''$ to the north-east from the centre,  which was later confirmed by \cite{moiseev} who also drew attention to the ``spur'' of HII regions in the north-east part of the galaxy which does not follow spiral arms. \cite{condon87} mapped the galaxy at 1.49 GHz and noted a continuum source $3'.5$ south of the galaxy. 

Observations and data reduction are presented in Section \ref{oa}, results of narrow band, broad band photometry and spectroscopy in Section \ref{results} of this paper. Section \ref{dis} discusses stellar populations with star formation rates and abundances of star forming regions. Conclusion is presented in Section \ref{con}.

\section{Observations and Analysis}
\label{oa}
\subsection{Optical imaging}
\label{imaging}
 
\begin{figure}
  \resizebox{\hsize}{!}{\includegraphics{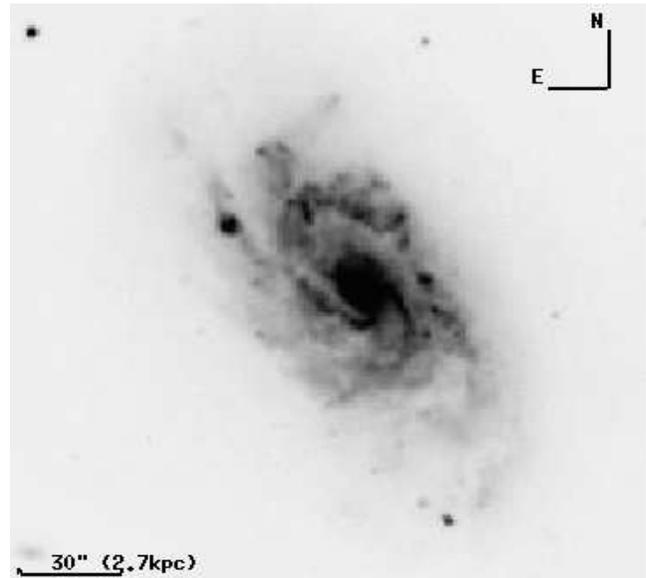}}
\caption[]{Galaxy NGC 1084 in $B$ band. North is up and East is towards left. Size of the image is $3'\times3'$ corresponding to $16.4\times16.4$ kpc$^2$. (H$_0$ = 75 kms$^{-1}${Mpc}$^{-1}$).}
\label{fig1}
\end{figure}

\begin{figure}
   \resizebox{\hsize}{!}{\includegraphics{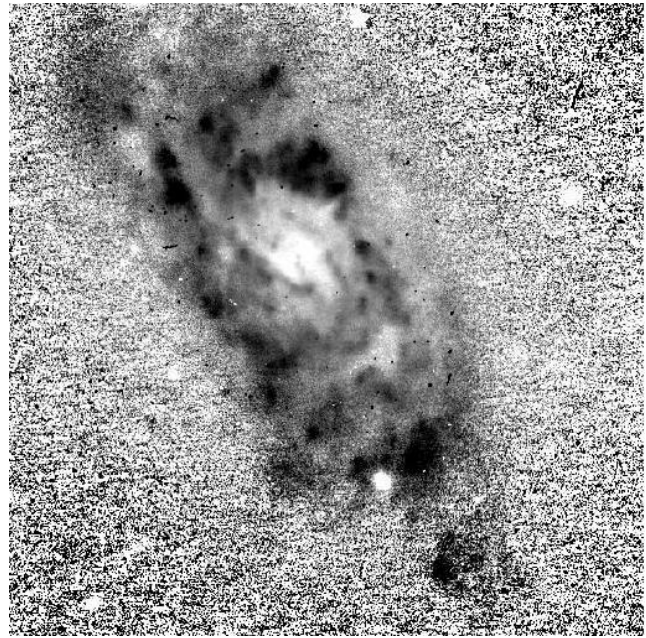}}
\caption[]{$B-V$ colour map of NGC 1084 reveals chaotic dust distribution. North is up and East is to the left. Dark and light shades in the image represent blue ($B-V$=0.19 mag) and red ($B-V$=0.83 mag) regions, respectively. Nucleus of the galaxy appears red due to large concentration of dust. A break in the spiral arm towards north-east  is due to the presence of a dust lane. Size of the image is $3'\times3'$ slightly offset to the south with respect to Figures \ref{fig1} and \ref{fig3}. The dust patch between the nucleus and bright star to the south is seen in Figures 1 and 2. In addition, in the southern part of the galaxy there is a blue spot below the bright star, continuing along the bridge to the dwarf galaxy (see Section \ref{dis}).}
\label{fig2}
\end{figure}

NGC 1084 was observed with the  2m Himalayan Chandra Telescope (HCT), Indian Astronomical Observatory, Hanle, India,   using the Himalaya Faint Object Spectrograph Camera (HFOSC), equipped with a 2K $\times$ 4K SITe CCD chip. The central  2K $\times$ 2K region, with a plate scale of  $0''.296/$pixel and the total field of view of 10 $\times$ 10 arcmin$^2$ was used for imaging. Observations were done under photometric sky condition on 04 December 2004. The log of imaging observations is given in Table \ref{tab1}. Images of the galaxy in Bessell's broad band  filters $U$$B$$V$$R$$I$   were obtained. 
Standard stars from the list of Landolt (1992) were observed to calibrate the broad band photometry.  The other calibration frames namely  bias frames were obtained throughout the night at an interval of about an hour; evening and morning twilight flats were also taken for flat field correction.  Data reductions were carried out using the standard packages available within IRAF \footnote{Image Reduction \& Analysis Facility Software distributed by National Optical Astronomy Observatories, which are operated by the Association of Universities for Research in Astronomy, Inc., under co-operative agreement with the National Science Foundation} software. Images were bias subtracted and  flat field corrected using the master bias and master twilight flat frames, respectively, to remove the pixel to pixel variation.  The images in different filters were aligned with respect to one another using  geometric mapping tasks {\it geomap}  and {\it geotran}, to bring them to the same co-ordinate system. The cosmic ray hits in the images were removed using the {\it cosmicrays} task. Sky background in the galaxy frame was estimated at the regions away from the galaxy and not affected by the stars. $B$ band image and $B-V$ colour map of the galaxy NGC 1084 are displayed in Figures \ref{fig1} and \ref{fig2}. The formal error introduced in the photometry by the calibration is of the order of 0.03-0.08 mag.\\

\begin{figure}
\resizebox{\hsize}{!}{\includegraphics{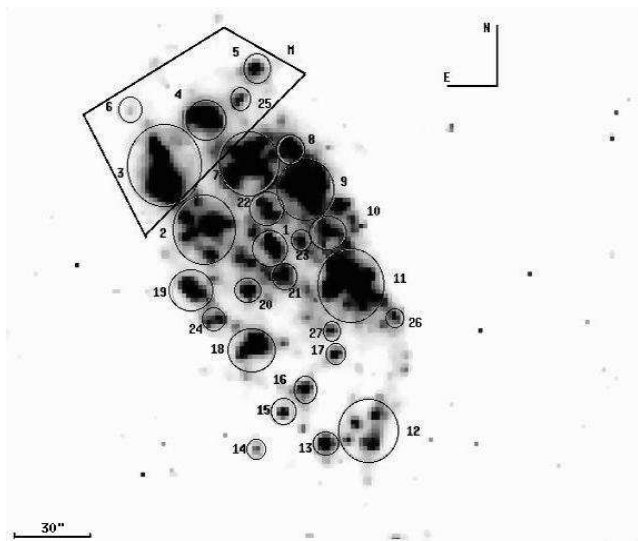}}
\caption[]{Continuum subtracted H$\alpha$ line image  of NGC 1084 reveals 27 star forming complexes spread over the face of the galaxy. The circles surrounding the star forming complexes represent the aperture used to get the flux of these complexes. Size of the image is $3'\times3'$. The trapezium labeled as M represents the ``spur'' region of \cite{moiseev}. See Section \ref{dis} for explanation.}
\label{fig3}
\end{figure}

The galaxy images were taken with  narrow band  H$\alpha$ (band width $\sim$100 \AA) filter.   For  flux calibration of narrow band H$\alpha$ image, spectrophotometric standard star Hiltner 102 was  observed. The pre-processing steps were similar to the broad band images.  Pure H$\alpha$ line image is obtained by subtracting a psf-matched and scaled version of the continuum image (which includes line emission also) from the H$\alpha$ image, as  described by \cite{waller}. $R$ band image of the galaxy has been used for continuum subtraction. As discussed by \cite{james}, for dark nights, the scaled $R$ band exposure gives an excellent continuum subtraction.  The continuum subtracted H$\alpha$ image is displayed in  Figure \ref{fig3}.

\begin{table}
\caption{Log of photometric observations.}
\label{tab1}
\centering
\begin{tabular}{cccc}\\
\hline{\bf Date of Obs.} & {\bf Filter} & {\bf Exposure} & {\bf Seeing } \\ 
& & (sec.)&\\
\hline\hline
04 Dec 2004 & $U$ & 960 & $2\farcs5$ \\
  & $B$ & 480 & $2\farcs0$ \\
  & $V$ & 240 & $1\farcs9$ \\
  & $R$ & 240 & $1\farcs8$ \\
  & $I$ & 240 & $1\farcs6$ \\
  & H$\alpha$ & 1800 & $1\farcs5$  \\
\hline
\end{tabular}
\end{table}

 
\subsection{Optical Spectroscopy}
Long-slit spectroscopic observations of the star forming complexes were done with a combination of grism Gr7 and 167l slit ($1\farcs92 \times 11'$), which covers the wavelength range $3800-6840 \ \rm{\AA}$ \ with a dispersion of 1.45 \AA/pixel (corresponds to 82 kms$^{-1}${pixel}$^{-1}$ at the central wavelength of 5320 \AA). To identify the complexes and to align the slit over these complexes, shorter exposures  in H$\alpha$ were taken. We have also determined the exact central coordinates of the complexes with the help of astrometry with respect to the USNO B1 stars registered in the frame.  Summary of spectroscopic observations is  given in Table \ref{tab2}.

\begin{figure}
  \resizebox{\hsize}{!}{\includegraphics{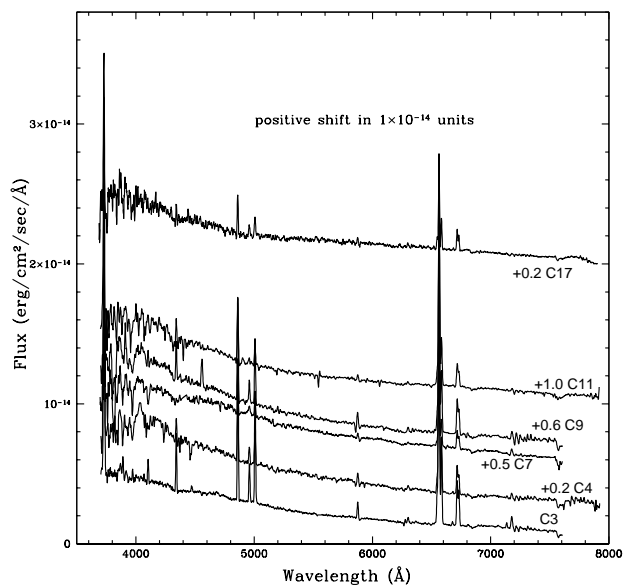}}
\caption[]{Spectra of a few of the bright complexes in NGC 1084,  these are displaced arbitrarily for clear viewing. The nomenclature follows Fig.3.}
\label{fig4}
\end{figure}

\begin{figure}
   \resizebox{\hsize}{!}{\includegraphics{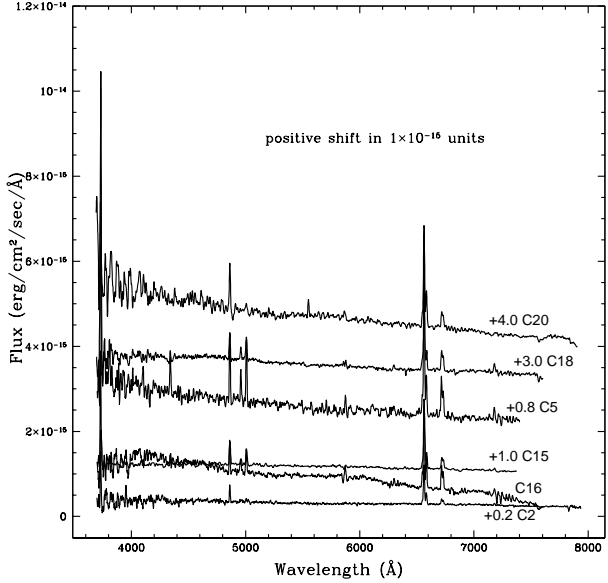}}
\caption[]{Same as the fig.4 for a few additional complexes.}
\label{fig5}
\end{figure}

\begin{table}
\caption{Log of spectroscopic observations. Complex in the second column refers to the star forming regions identified in Fig \ref{fig3} (see Section \ref{imaging} \& \ref{narrowbandimaging}).}
\label{tab2}
\centering
\begin{tabular}{ccc}\\
\hline{\bf Date of Obs.} & {\bf Complex} & {\bf Exposure} \\ \\
 
\hline\hline
 
05 Dec 2004 &  C1 & 900 s \\
 & C3 & 900 s \\
12 Dec 2004 & C2 & 900 s \\
 & C7 & 900 s \\
 & C18 & 1800 s \\
13 Dec 2004 & C4 & 1800 s \\
 & C9 & 1800 s \\
 & C11 & 900 s \\
05 Sep 2005 & C15 & 900 s \\
 & C16 & 900 s \\
 & C17 & 2700 s \\
 & C20 & 1800 s \\
06 Sep 2005 & C5 & 900 s \\
 
\hline
 
\end{tabular}
\end{table}
Spectroscopic reduction involves extraction of the spectra using {\it apall} task in {\it specred} package of IRAF. Wavelength calibration of the spectra is achieved using the  Ferrous-Argon lamp spectrum. The combination of  grism Gr7 and 167l slit gives a spectral resolution of $\sim$10 \AA. Spectrophotometric standard Hiltner 600 was used for flux calibration of the spectra.  The reduced spectra of the knots have been plotted in Figures \ref{fig4} and  \ref{fig5}.


\section{Results}
\label{results}
\subsection{Broad band imaging}
$B$ and $B-V$  images of the galaxy are displayed in Figures \ref{fig1} and \ref{fig2}.  In the broad band  image, the spiral arms, knotty structures within the arms  and  intermittent prominent dust lanes are clearly seen.   The main body of the galaxy seems to have very faint extension in the north-east and south-west direction, with many faint knots embedded in them.  The $B-V$ colour image of the galaxy can be used to trace the  distribution of the star forming knots and dust in the galaxy. The nucleus of the galaxy has very red colour ($B-V=0.83 \ \rm{mag}$), indicative of large amount of dust concentration, surrounded by spiral arms, forming a blue ring ($B-V=0.3 \ \rm{mag}$). The extent of the dust patch at the nucleus as measured from the $B-V$ colour map is $\sim$1 kpc. The faint extensions seen in the broad band images are blue in colour, indicative of ongoing star formation in these regions too. Comparison of $B-V$ colour image with the pure H$\alpha$ emission line image (refer Figure \ref{fig3}) shows that the bluer regions in the $B-V$ colour image correspond to the prominent emission regions in the H$\alpha$ line image, though the sizes of both the regions are not identical, as noted by \cite{elmegreen} in a sample of spiral and irregular galaxies.

The  broad band $U$$B$$V$$R$$I$ photometry of the complexes is  presented in Table \ref{tab3}. To get accurate photometry, a proper choice of the source and background aperture sizes is important. One of the main source of  error in photometry is the crowding of the complexes. It can be seen from the continuum image (Figure \ref{fig1}) and continuum subtracted H$\alpha$ image (Figure \ref{fig3}) that the star forming complexes have different shapes and sizes, which indicates the need to  define  apertures and background regions for individual complexes. Centres of the complexes were estimated from the continuum subtracted H$\alpha$ image, and same centres were used to obtain the broad band magnitudes using aperture photometry with the aperture radii overplotted on the continuum subtracted H$\alpha$ image in Figure \ref{fig3} and listed in Table \ref{tab3}. Galaxy background subtraction plays a crucial role in the photometry of star forming  complexes.  We have estimated  the galaxy background in an annulus just beyond the integration aperture of each  complex.
 Photometric calibration of the data is done using \cite{landolt} standards. The reddening correction for different complexes are determined as discussed in Section 3.3.2. The dereddened colours for the complexes are also presented in Table \ref{tab3}.
\begin{table*}
  \begin{minipage}{126mm}
\caption{Reddening corrected magnitudes and colours for each of the complexes.}
\label{tab3}
\centering
\begin{tabular}{cccccccc}
  \hline {\bf ID} & {\bf Aperture} & ${\bf E(B-V) }$ & ${\bf V}$ & ${\bf(U-B)_0}$ & ${\bf(B-V)_0}$ & ${\bf(V-R)_0}$ & ${\bf(V-I)_0}$\\ 

& arcsec & & & & & &  \\

  \hline\hline
1 & 5.4    & 0.45 &  13.409 & -0.170 & 0.372 & 0.204 & 0.624 \\
2 & 9.5    & 0.32 &  13.723 &  2.047 & 0.593 & 0.255 & 0.558 \\
3 & 12.0   & 0.32 &  13.279 & -0.813 & 0.078 & 0.132 & 0.221 \\
4 & 6.0    & 0.45 &  14.453 & -0.903 &-0.075 & 0.093 & 0.053 \\
5 & 4.8    & 0.07 &  17.397 & -0.875 & 0.262 & 0.353 & 0.665 \\
6 & 3.6    & 0.32 &  17.937 & -1.049 & 0.284 & 0.307 & 0.519 \\
7 & 9.0    & 0.32 &  14.786 & -1.085 &-0.486 & -0.034 & -0.685 \\
8 & 4.2    & 0.45 &  14.834 & -1.011 &-0.152 & 0.023 & -0.018 \\
9 & 9.0    & 0.45 &  14.513 & -1.708 & 0.265 & 0.385 & 1.230 \\
10 & 4.8   & 0.45 &  16.289 & -1.493 & 0.797 & -0.124 & 0.140 \\
11 & 10.2  & 0.45 &  12.875 & -0.604 &-0.060 & 0.057 & 0.082 \\
12 & 9.0   & 0.15 &  16.640 & -1.156 &-0.144 & 0.500 & 0.477 \\
13 & 3.6   & 0.15 & 19.019 & -1.786 & 0.031 & 0.753 & 0.997 \\
14 & 2.4   & 0.15 & 20.218 & -1.569 &-0.168 & 0.434 & 0.285 \\
15 & 3.6   & 0.15 & 18.355 & -0.813 & 0.063 & 0.338 & 0.108 \\
16 & 3.6   & 0.15 & 18.386 & -0.817 & 0.699 & 0.415 & 0.751 \\
17 & 3.0   & 0.45 & 18.847 &  0.184 &-0.280 & 0.665 & 0.483 \\
18 & 7.2   & 0.07 & 15.318 & -0.452 & 0.386 & 0.347 & 0.708 \\
19 & 6.0   & 0.07 & 15.866 &  0.586 & 0.545 & 0.291 & 0.640 \\
20 & 3.6   & 0.45 & 16.505 & -0.248 &-0.165 & -0.001 & 0.281 \\
21 & 3.6   & 0.45 & 14.942 & -0.330 & 0.284 & 0.171 & 0.420 \\
22 & 4.8   & 0.45 & 14.969 & -0.155 & 0.340 & 0.364 & 0.741 \\
23 & 2.4   & 0.45 & 15.755 & -0.359 & 0.504 & 0.172 & 0.319 \\
24 & 4.5   & 0.07 & 16.839 & -0.795 & 0.316 & 0.287 & 0.617 \\
25 & 2.4   & 0.45 & 17.847 & -1.025 & 0.143 & 0.117 & 0.121 \\
26 & 2.4   & 0.45 & 16.497 & -0.878 & 0.041 & 0.010 & 0.150 \\
27 & 2.1   & 0.45 & 17.876 & -1.264 & 0.157 & 0.423 & 0.605 \\
\hline
\end{tabular}
\end{minipage}
\end{table*}\\
\subsection{Narrow band imaging}
\label{narrowbandimaging}
 As can be seen from the continuum subtracted image (Figure \ref{fig3}) of the galaxy, the star forming regions are not well isolated, instead they are  extended and  distorted and hence make it difficult to define the border of individual region. Many HII regions are overlapping in the image and one will undoubtedly miss a number of HII regions that are too weak to be detected in the vicinity of stronger emitters close by \citep{rozas}.

Objective criteria have been used in the literature to define the sizes of HII regions with the aim of determining their luminosity functions (\citealt{bradley06} and \citealt{fathi07}). \cite{rand} selected the HII regions based on the criteria that they are centrally peaked with the peak flux exceeding the local background by a factor of 1.5. \cite{rozas} selected HII regions only if they comprise of at least nine continuous pixels, each with an intensity of at least 3 times the rms noise level of the local background. Here we apply a similar criterion that the HII regions are centrally peaked with peak flux having an intensity of at least 5 times the rms noise level of the local background. Since our intention is to obtain accurate H$\alpha$ as well as continuum fluxes for selected regions, we have chosen aperture sizes such that broad band $U$$B$$V$$R$$I$ magnitudes are accurate to 0.1 mag. This has resulted in clubbing more than one HII region in a single aperture when the underlying continuum is faint.

 In this way we group the star forming regions in this galaxy into 27 complexes (Figure \ref{fig3}). \cite{haynes99} and \cite{kori04} using HI data calculated the radial velocity of the galaxy as 1411 kms$^{-1}$ and 1400 kms$^{-1}$, assuming velocity = 1407 kms$^{-1}$ (NED) and H$_0$ = 75 kms$^{-1}${Mpc}$^{-1}$ the distance to NGC 1084 is calculated as 18.76 Mpc ($1''$ corresponds to 91 pc at this distance). The extent of these luminous HII regions varies from a few hundred pc to a few kpc. Almost all the Galactic HII regions have diameters much less than 50 pc. The characteristic diameter obtained by \cite{vanden81} for our galaxy is around 44 pc and the minimum characteristic diameter of HII regions obtained for other nearby spiral galaxies was 25 pc. However, \cite{vanden81} and \cite{hodge87}, in their study concluded that the mean characteristic diameter of HII regions for spirals is around 150 pc. The minimum size of the complex obtained for NGC 1084 is 191 pc for complex C27 and the maximum size obtained is around 1.1 kpc for complex C3. Thus, the star forming complexes identified by us in the continuum subtracted H$\alpha$ image of NGC 1084 are groups of HII regions, ionized by multiple OB star clusters.

The star formation history of galaxies can be traced using the integrated H$\alpha$ equivalent widths, which are usually measured spectroscopically, however,  the continuum subtracted H$\alpha$ image of the galaxy can be used more effectively to include emission from the entire complex. One of the main problems in deriving star formation related quantities using narrow band H$\alpha$ imaging is the need for large and uncertain extinction corrections, and correction for the contribution due to [NII]  emission. The spectroscopically measured Balmer decrement is a good tracer of the extinction within the star forming region. Further, the contamination of H$\alpha$ flux and equivalent widths due to  [NII] lines at $\lambda$6548 \AA \ and $\lambda$6584 \AA \ is found to vary from galaxy to galaxy \citep{kenni92} and within a galaxy \citep{green}. 
It is  concluded by \cite{dopita02} that the H$\alpha$ imaging can be used as a reliable indicator of star formation in starburst galaxies only if the spectroscopic data is also available to correct for internal extinction and contribution due to [NII]  emission. As a first order estimate [NII] emission is taken to be 30\% of the total H$\alpha$+[NII] flux as given by \cite{kenni83}.  
The  flux in the star forming complexes have been measured from the continuum subtracted H$\alpha$ line image, within the aperture listed in Table 3 and corrected for  contributions due to reddening measured from the spectroscopic Balmer decrement (see Section \ref{sec:red}) and due to \ [NII]. The  corrected H$\alpha$ fluxes  along with the H$\alpha$ luminosity for the complexes have been given in Table \ref{tab4}. The H$\alpha$ equivalent widths of the complexes estimated using the expression  given by \cite{waller} are shown in Table \ref{tab5}. Strong H$\alpha$ emission from the complexes is the  evidence for a very young ($< 8$ Myr) population of massive stars \citep{whitmore}.

\begin{table*}
\begin{minipage}{126mm}
\caption{Positions and H$\alpha$ luminosities of the star forming complexes.}
\label{tab4}
\begin{center}

\begin{tabular}{cccccc}
\hline {\bf ID} & ${\bf RA}$ & ${\bf DEC}$ & ${\bf log[f(H\alpha+[NII]]}^1$ & ${\bf log[f(H\alpha)]}^2$ & ${\bf log(luminosity)}$  \\
\\
 & hh:mm:ss.ss & dd:mm:ss.s & & & \\

\hline\hline

 1 & 02 45 59.92 & -07 34 42.6 & -12.687 & -12.384 & 40.240 \\
 2 & 02 46 01.24 & -07 34 38.0 & -12.265 & -12.093 & 40.531 \\
 3 & 02 46 02.05 & -07 34 19.6 & -12.009 & -11.838 & 40.786 \\
 4 & 02 46 01.24 & -07 34 07.5 & -12.590 & -12.286 & 40.338 \\ 
 5 & 02 46 00.24 & -07 33 53.6 & -13.121 & -13.205 & 39.420 \\
 6 & 02 46 02.68 & -07 34 05.0 & -13.824 & -13.653 & 38.971 \\
 7 & 02 46 00.43 & -07 34 19.4 & -12.197 & -12.026 & 40.599  \\
 8 & 02 45 59.57 & -07 34 15.9 & -12.784 & -12.481 & 40.143 \\
 9 & 02 45 59.28 & -07 34 26.6 & -12.104 & -11.800 & 40.824 \\
10 & 02 45 58.86 & -07 34 38.2 & -12.901 & -12.597 & 40.027 \\
11 & 02 45 58.43 & -07 34 53.7 & -12.161 & -11.857 & 40.767 \\
12 & 02 45 58.09 & -07 35 33.0 & -12.678 & -12.681 & 39.944 \\
13 & 02 45 58.95 & -07 35 36.5 & -13.179 & -13.182 & 39.443 \\
14 & 02 46 00.28 & -07 35 38.1 & -13.886 & -13.889 & 38.736 \\
15 & 02 45 59.76 & -07 35 28.0 & -13.448 & -13.450 & 39.175 \\
16 & 02 45 59.32 & -07 35 21.5 & -13.248 & -13.250 & 39.374  \\
17 & 02 45 58.76 & -07 35 12.1 & -13.444 & -13.140 & 39.484 \\
18 & 02 46 00.37 & -07 35 09.9 & -12.581 & -12.664 & 39.960 \\
19 & 02 46 01.52 & -07 34 54.3 & -12.776 & -12.860 & 39.764 \\
20 & 02 46 00.39 & -07 34 54.7 & -13.178 & -12.874 & 39.750 \\
21 & 02 45 59.70 & -07 34 50.0 & -13.098 & -12.794 & 40.830 \\
22 & 02 46 00.02 & -07 34 32.5 & -12.919 & -12.615 & 40.009 \\
23 & 02 45 59.38 & -07 34 41.1 & -13.454 & -13.151 & 39.473 \\
24 & 02 46 01.12 & -07 35 03.1 & -13.159 & -13.243 & 39.382 \\
25 & 02 46 00.57 & -07 34 02.7 & -13.582 & -13.278 & 39.346 \\
26 & 02 45 58.83 & -07 35 05.9 & -13.615 & -13.311 & 39.313 \\
27 & 02 45 57.58 & -07 35 02.0 & -13.668 & -13.364 & 39.260 \\
\hline
\end{tabular} \\
\end{center}
$^1$ Observed flux.\\
$^2$ H$\alpha$ flux corrected for reddening and contribution due to [NII].

\end{minipage}
\end{table*}

\subsection {Spectroscopy}
\subsubsection{Emission line fluxes}
The integrated fluxes of the main nebular lines H$\alpha$, H$\beta$, [OII] $\lambda$3727 \AA \ and [OIII] $\lambda$4959,5007 \AA \ have been measured from  the reduced  spectra of the star forming complexes. The measured Balmer line fluxes of star forming regions are always an underestimate of the real flux due to the presence of underlying stellar absorption of hydrogen lines and also the internal extinction due to the dust present in these regions. \\
 The absorption equivalent width due to the underlying stellar population depends on the age of starburst and star formation history (\citealt{izotov}, \citealt{gonz}) which is uncertain and hence we adopt a constant absorption equivalent width of 2 \AA \ for all the hydrogen lines (\citealt{pop}, \citealt{mccall85}, \citealt{searle78}). \cite{mccall85}  assumed that the first four Balmer series H$\alpha$, H$\beta$, H$\gamma$ and H$\delta$ all have the same absorption equivalent widths similar to the assumption of \cite{searle78} and used H$\gamma$ and H$\beta$ equivalent widths to determine the absorption equivalent width of underlying Balmer absorption. The above assumption is valid as the model atmospheres by \cite{kurucz79} show that the absorption equivalent widths for H$\alpha$, H$\beta$ and H$\gamma$ agree to within 30\% and H$\beta$ and H$\gamma$ equivalent widths agree to within 3\% for any given star \citep{mccall85}.
\cite{searle78} and \cite{mccall85} tried to obtain a best fit to the theoretical Balmer decrement H$\alpha/$H$\beta$ from the observed  H$\alpha/$H$\beta$ fluxes for which reddening corrections have already been applied, and estimated absorption equivalent widths for 3 HII regions in M101 to be 2 \AA \  and  1.9 \AA, \ respectively. Here, we adopt this value of 2 \AA \ for correcting for the underlying stellar population using  Eq. 3 of \cite{kong}
\begin{center}
\ensuremath{{{F_{line}}^{cor}} = {{F_{line}}^{obs}}(1 + {{EW_{line}}^{abs}}/{{EW_{line}}^{obs}})}
\end{center}
where ${{F_{line}}^{cor}}$ and ${{F_{line}}^{obs}}$ are the absorption corrected and observed emission line fluxes, respectively.  ${{EW_{line}}^{obs}}$  is the observed equivalent width of the emission lines and ${{EW_{line}}^{abs}}$ is the equivalent width of underlying stellar absorption taken as 2 \AA.\\

The 2 \AA \ absorption equivalent width is a factor of 3 lower than would be predicted for a normal unevolved OB association. This fact was explained by \cite{mccall85} as due to the enhancement of the upper IMF which decreases absorption equivalent widths due to the contribution by the high mass stars of significant continuum flux and whose fractional population increases with the metal abundance. The second alternative explanation for lower value of absorption equivalent width could be that, the OB association might contain evolved stars, supergiant OB stars contributing to the observed continuum flux and thereby diluting the absorption equivalent width to a value $\sim$2-3 \AA.

Stellar absorption correction is crucial as the underlying Balmer absorption lines steepen the emission Balmer decrement H$\alpha/$H$\beta$ significantly which leads to an overestimate of the extinction which in turn influences the other line ratios. This we tried to check in our spectra while calculating the $E(B-V)$ values (see Section \ref{sec:red}) for various complexes for which spectra were obtained. $E(B-V)$ value for complex C7 was found to be 0.76 mag when the stellar absorption correction was not applied and 0.32 mag after the 2 \AA \ absorption equivalent width correction. A decrease of $\sim$0.2 mag in $E(B-V)$ was observed for several other complexes.

\subsubsection{Reddening} 
\label{sec:red}
 The  reddening $E(B-V)$ in the Milky Way in the direction of  NGC 1084  as determined by \cite{burstein} using the HI column density  is 0.013 mag while another estimate by \cite{schlegel} using the COBE and IRAS maps is 0.027 mag. We have taken $E(B-V)$ = 0.027 mag to correct for the Galactic reddening.

A comparison between   pure H$\alpha$ image and $B-V$ colour image of NGC 1084 reveals that there are several star forming complexes e.g. C1, C9, C10, C11, C21, C22, C23 etc. which are heavily extincted by the dust, whereas there are other complexes which are relatively less obscured, indicating a non-uniform distribution of the dust. In such cases  determination of internal reddening within the galaxy is very important. This can be achieved either by classifying the star forming complexes into different extinction bins, according to how much dust is present in the immediate vicinity of the  cluster \citep{hunter}, with a priori knowledge of the variation of the reddening within the galaxy by some other means, or by estimating  the reddening for the individual star forming complexes by spectroscopic observations. Here, we have opted  for an intermediate approach by determining $E(B-V)$ spectroscopically for some brighter star forming complexes namely C1, C2, C3, C4, C5, C7, C9, C11, C15, C16, C17, C18 and C20, spread over the face of the galaxy and assigning the same reddening to the other nearby complexes. 

The internal extinction due to dust can be estimated by comparing the Balmer line ratio H$\alpha$/H$\beta$ obtained after correcting for the underlying stellar absorption to the theoretical Balmer decrement (2.87) for case B recombination at 10$^{4}$ K.  The internal  colour excess ${E_{B-V}}^{int}$ is calculated using the relation given by \cite{kong}.

\begin{center}
\begin{equation}
\ensuremath{{E_{B-V}}^{int} = A_v/R_v = 1.086{\tau}_v/R_v}  
\end{equation}
\end{center}
where,\\
\begin{center}
\ensuremath{ {\tau}_v = -{\frac{ln[F_{H\alpha}/F_{H\beta}] - ln[I_{H\alpha}/I_{H\beta}]}{({{\lambda}_{H\alpha}/5500})^{-0.7} - ({{\lambda}_{H\beta}/5500})^{-0.7}}}}
\end{center}
based on the effective absorption curve \ensuremath{{\tau}_{\lambda} = {\tau}_v(\lambda/5500)^{-0.7}} given by \cite{charlot}. \ensuremath{[F_{H\alpha}/F_{H\beta}]} is the ratio of stellar absorption corrected observed line fluxes of H$\alpha$ and H$\beta$.
\ensuremath{[I_{H\alpha}/I_{H\beta}] = 2.87} for starburst galaxies.  The spectroscopically determined reddening values for the complexes are listed in Table \ref{tab6}.  To get an estimate of reddening for the other complexes we assigned $E(B-V)$ values depending on its proximity to  the complex for which we have spectroscopically determined $E(B-V)$ value. In this way all the complexes can be put in extinction bins. Complexes C1, C4, C8, C9, C10, C11, C17, C20, C21, C22, C23, C25, C26 and C27 which are located in the inner region of the galaxy, having very red colour in the $B-V$ image, indicative of presence of heavy internal extinction,  have been assigned $E(B-V)_{tot}$ =$ 0.45$. Complexes C2, C3, C6 and  C7 which lie on the north-east side of the nucleus can be put in the second bin with an intermediate value of $E(B-V)_{tot} = 0.32$. In the third bin with $E(B-V)_{tot} =0.15$ complexes C12, C13, C14, C15 and C16 can be grouped, which are lying on the southern part of the galaxy. Remaining complexes namely C5, C18, C19 and C24 are considered to have lowest value of $E(B-V)_{tot}$ = $0.07$. Our estimates of the internal reddening are similar to those of \cite{ho97} for the HII regions in the centres of  nearby galaxies. 

We would like to mention here that the observed F$_{H\alpha}$/F$_{H\beta}$  ratio for complex C18 is less than the assumed theoretical ratio of 2.87, moreover, the \( \frac{[\rm NII] \lambda 6548}{[\rm NII]\lambda\lambda 6548,6584 + H\alpha} \) ratio is about 16\% unlike in the other complexes where it is found to be around  5-6\%. Assuming  5-6\% as the standard ratio, excess of 11\% of the $[\rm NII] \ \lambda6548 \ \AA$ flux is added back to the H$\alpha$ flux. Now the ratio of observed \ F$_{H\alpha}$/F$_{H\beta}$ \ for complex C18, after the stellar absorption correction is = 2.99, a little more than the theoretical 2.87 value, which translates to  the colour excess ${E_{B-V}}^{int}$=0.07 as shown in Table \ref{tab6}. From the reddening distribution on the face of the galaxy it is clear that the complexes C8, C9, C10, C11  have the highest reddening and complexes C18, C19 and C24 have the least reddening, it indicates that complexes C18, C19 and C24 which are in the south-east part of the galaxy, are towards the observer, whereas complexes C8, C9, C10, C11 occupy the region which is away from the observer; similar conclusion was also arrived at by \cite{moiseev}.

\section{Discussion}
\label{dis}
\subsection{Colour-colour diagrams}
\label{ref:c-c}

\begin{figure}
\resizebox{\hsize}{!}{\includegraphics{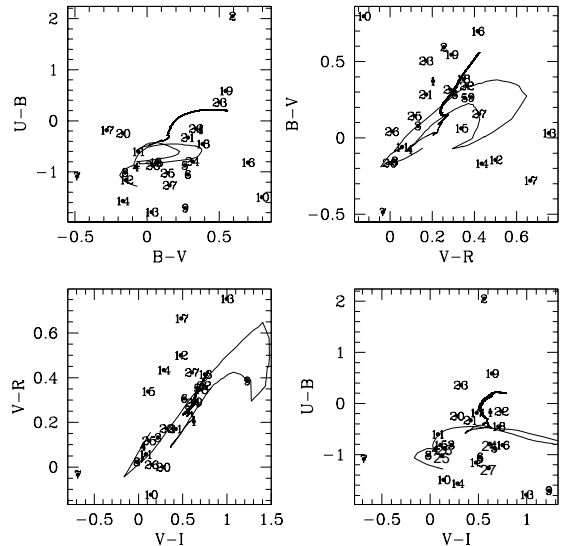}}
\caption[]{Colour-Colour diagram of the complexes. Tracks are the SB99 model tracks with Salpeter IMF and solar metallicity.}
\label{fig7}
\end{figure}

Figure \ref{fig7} shows the colour-colour diagram for the complexes. Starburst99 \citep{leith99} model tracks are shown for solar metallicity with Salpeter IMF for mass ranging from 1 $M_{\odot}$ to 100 $M_{\odot}$. The complexes do not fall exactly on the tracks even after correcting for internal extinction.  The deviation of complexes C2, C13, C17, C10, C23, C19 from the superimposed track may be due to the fainter detection limit in H$\alpha$ than in the broad band filters, which resulted in more uncertainty in the colours. Also, a few complexes may have chaotic dust distribution which cannot be corrected completely with simple extinction correction. It has been observed by \cite{weistrop} and \cite{mayya}  that the continuum extinctions are about twice greater than the ones determined from spectroscopy and hence needs rigorous modelling \citep{calzetti94}. Another possible explanation could be that these complexes have multiple populations embedded in them which is displacing their position away from the tracks.

\begin{figure}
\resizebox{\hsize}{!}{\includegraphics{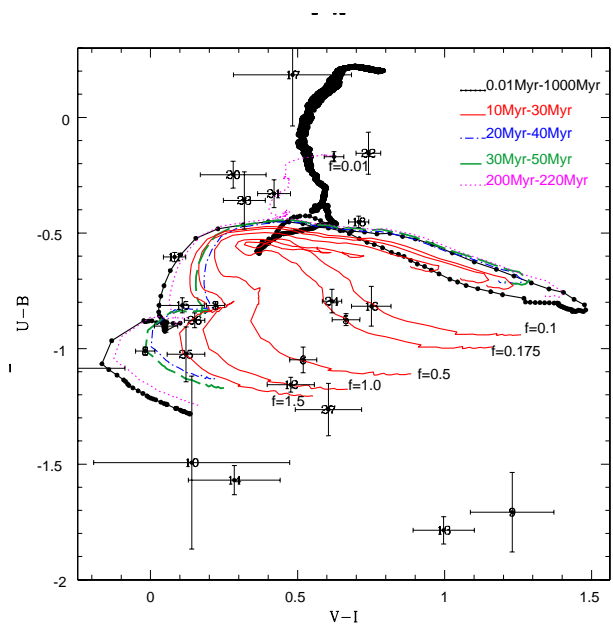}}
\caption[]{$U-B$ vs $V-I$ composite colour-colour diagram created for mixed populations. Solid curve represents the mix of 0.01-10 Myr recent young ionizing component with the 10-20 Myr old component along with different proportions which is given by the fraction $f$=1.5, 1.0, 0.5, 0.175, 0.1. $f$ is the ratio of young component to the old component. Dash-dotted curve corresponds to a composite mix of 0.01-10 Myr with 20-30 Myr population. Dashed curve corresponds to mixed populations of 0.01-10 Myr and 30-40 Myr. Dotted curves correspond to young component 0.01-10 Myr mixed with 200-210 Myr old component, mixed at different proportions. $f$=1.0 and 0.01. Thick black line with black dots corresponds to original SB99 colour-colour curve with single age plotted from 0.01-1000 Myr. See Section \ref{ref:c-c} for explanation. Error bars here refers to the photometric errors.}
\label{fig8}
\end{figure}

We try to investigate the possibility of multiple populations using models created from the Starburst99 model tracks. The Starburst99 model evolutionary tracks offer a range of initial conditions namely initial mass function (IMF), metallicity and mode of star formation i.e. continuous and instantaneous. The oxygen abundance  for some bright  knots were determined using medium resolution spectroscopy (Section 4.3) and it is found that  the oxygen abundance is close to solar. We checked amongst the three possible IMF available with Starburst99 model and found it difficult to constrain the IMF using $(U-B)$ and $(B-V)$ colours. So, we have adopted a Salpeter IMF with lower and upper mass limits of 1 $M_\odot$ and 100 $M_\odot$, which are commonly used for star forming regions. Out of the two possible models of star formation we have selected instantaneous burst as the star forming regions in our sample are  similar to the star clusters or associations \citep{weistrop}. So the mixed population model is created from Starburst99 undergoing an instantaneous burst with Salpeter IMF in the mass range 1 $M_\odot$ to 100 $M_\odot$. We consider the recent burst to have an age between 0.01 Myr and 10 Myr and add a population 10 Myr older than this. Similarly we construct evolutionary tracks for current burst of age 0.01-10 Myr mixed with an older population seperated by 20, 30 and 200 Myr. For each age range including the young component, the corresponding $M_U$, $M_B$, $M_V$ and $M_I$ were read off from the Starburst99, Salpeter IMF models.  $M_U$ of 0.01-10 Myr was mixed with each of 10-20 Myr, 20-30 Myr, 30-40 Myr and 200-210 Myr in the flux domain and a composite $M_U$ was obtained. Similarly $M_B$, $M_V$ and $M_I$ were mixed in the flux domain.  These magnitudes were used to calculate composite $U-B$ and $V-I$ colours for various mix. The mixed population curves are plotted as $U-B$ vs $V-I$ colour-colour diagram in Figure \ref{fig8}. $R$ band is not considered here because of contribution due to the H$\alpha$ emission.

The solid curve in Figure \ref{fig8} corresponds to a mix of 0.01 Myr and 10 Myr populations evolving over 10 Myr. Dash-dotted, dashed and dotted curves represent evolution of a mix of 0.01 Myr with 20 Myr, 30 Myr and 200 Myr population, and evolved for 10 Myr. Thick black line with black dots corresponds to original, single age Starburst99 model evolutionary track plotted from 0.01-1000 Myr for reference. Observed colours of many of the star forming complexes namely C3, C4, C8, C11, C12, C15 and C18 fit into these curves, giving us an idea about the plausible age range of the old population embedded in them. These curves trace a mix of equal proportion of young to old population. Next we consider the mix of young and old population in different proportions $f$, where $f$ is the ratio of young to old component. Increasing contribution of young component ($f>$1.0) was considered by \cite{diaz} in their study of HII regions of four different galaxies. We obtain model curves for $f$=0.1, 0.175 and 0.5 representing a young population of 10\%, 17.5\% and 50\% of the older population, and f=1.5 representing the young population 1.5 times stronger than the old population.

As seen in Figure \ref{fig8}, complex C16 lies on the $f$=0.1 curve indicating that the strength of the recent burst is only 10\% of the older burst that took place 10 Myr earlier. Similarly C5, C6, C12, C24, C25, C26, C27 fall on curves with different strengths of the current burst with respect to the burst that took place 10 Myr earlier. The complexes C3 and C15 clearly have an underlying component 20 Myr older than the current one, and it is likely that C8 may also have similarly older component. The underlying component in complexes C4, C10, C11 and C18 appear much older. It is interesting that the complexes C1 and C21 can be modelled with fairly evolved recent burst which is at 1\% level of a 200 Myr old burst, and C17, C20, C22, C23 may also be modelled with small variations in the model. It thus appears possible that recurrent starbursts have taken place in the circumnuclear region about 200 Myr ago, and have taken place in the disk more recently. \cite{allard06} had reached similar conclusion on the circumnuclear ring of M100, that the star formation had occurred in a series of short bursts over the last 500 Myr triggered by the feeding of gas from the disk to the bar. 

\begin{figure}
\resizebox{\hsize}{!}{\includegraphics{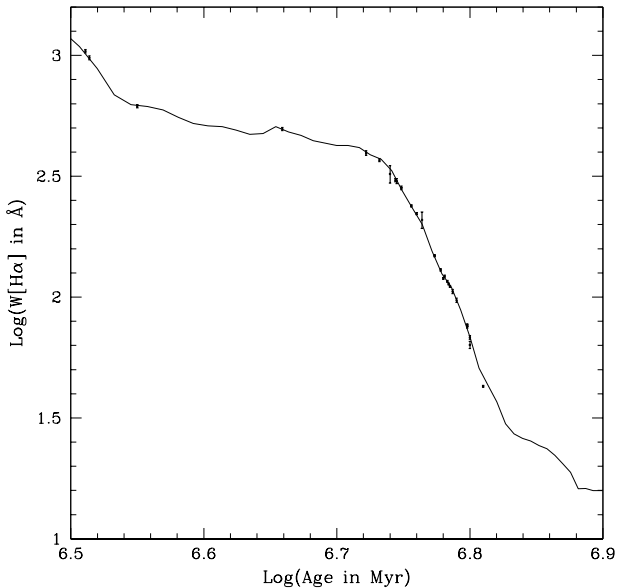}}
\caption[]{Plot showing the equivalent width of H$\alpha$  versus age of the complexes. Model track is taken from Starburst99 model. Ages estimated for the complexes are between 3.2 Myr to 6.45 Myr. Errorbars plotted here correspond to error on the equivalent width.}
\label{fig9}
\end{figure}

\subsection{Age of the complexes}
Because of the degeneracy between age and extinction, both producing a reddening, the age cannot be determined in a unique way from the colours. The difficulty is compounded by the presence of older bursts. However, the equivalent width of H$\alpha$ can be used to estimate the age,  as it decreases with age. Therefore, ages of the complexes are determined by comparing the equivalent width of the H$\alpha$ emission line, measured from the continuum subtracted H$\alpha$ image (Figure \ref{fig3}) of the galaxy with the Starburst99 model prediction and are listed in Table \ref{tab5}.  The spread in the age of the complexes is 3.24 to 6.45 Myr. We do not expect to see  star forming regions with age less than 3 Myr, as they are buried deep in the molecular clouds, and greater than 7 Myr since the ionizing flux becomes too low to produce detectable H$\alpha$ emission. 

While a spectrum of OB stars distributed in an initial mass function ionize the star forming complexes, often the number of equivalent O5 stars is used as a measure of the size of the young cluster.  The number of Lyman continuum photons emitted by an O5 star is of the order of ${10}^{49}$ photons s$^{-1}$  and log$(N_{lym})=52.4$ implies a total of 1000 O5 stars \citep{ravin}. Table 5 gives the total number of equivalent O5 stars in the HII complexes of NGC 1084.

\begin{table*}
\begin{center}
\caption{Table showing H$\alpha$ equivalent width, H${\alpha}$ Luminosity, No. of Lyman photons, the star formation rate and age calculated using the Starburst99 model, for all the complexes. Refer Figure \ref{fig3} for identification} 
\label{tab5}
\begin{tabular}{ccccccc}
  \hline {\bf ID}  & ${\bf EW(H\alpha)}$ & ${\bf log(Lum.)}$ & ${\bf log(N_{lym})}$ & ${\bf SFR}$ &  ${\bf AGE}$ & ${\bf No. of \ O }$ \\
 & $\AA$ & & & $\rm{M_\odot}{yr}^{-1}$ & $Myr$& {\bf stars}\\

\hline \hline
1  & 37.25 & 40.24 & 52.11 & 0.14 & 6.45 & 513\\
2  & 97.06 & 40.53 & 52.40 & 0.27 & 6.16 & 1000\\
3  & 116.2 & 40.79 & 52.65 & 0.48 & 6.07 & 1778\\
4  & 122.1 & 40.34 & 52.20 & 0.17 & 6.04 & 631\\
5  & 221.6 & 39.42 & 51.29 & 0.02 & 5.75 & 78\\
6  & 129.6 & 38.97 & 50.84 & 0.01 & 5.99 & 28\\
7  & 302.1 & 40.60 & 52.46 & 0.31 & 5.55 & 1148\\
8  & 110.7 & 40.14 & 52.01 & 0.11 & 6.09 & 407\\
9  & 394.5 & 40.82 & 52.69 & 0.53 & 5.27 & 1950\\
10 & 323.2  & 40.03 & 51.89 & 0.08 & 5.49 & 309\\
11 & 76.59  & 40.77 & 52.63 & 0.46 & 6.28 & 1698\\
12 & 368.8  & 39.94 & 51.81 & 0.07 & 5.39 & 257\\
13 & 1041.0  & 39.44 & 51.31 & 0.02 & 3.24 & 81\\
14 & 616.6  & 38.74 & 50.60 & 0.00 & 3.55 & 16\\
15 & 304.6  & 39.17 & 51.04 & 0.01 & 5.54 & 44\\
16 & 496.0  & 39.37 & 51.24 & 0.02 & 4.56 & 69\\
17 & 977.5  & 39.48 & 51.35 & 0.02 & 3.26 & 89\\
18 & 113.3  & 39.96 & 51.83 & 0.07 & 6.08 & 269\\
19 & 119.6  & 39.76 & 51.63 & 0.05 & 6.02 & 170\\
20 & 208.6  & 39.75 & 51.62 & 0.04 & 5.81 & 166\\
21 & 59.39  & 39.83 & 51.70 & 0.05 & 6.31 & 200\\
22 & 92.05  & 40.01 & 51.87 & 0.08 & 6.12 & 295\\
23 & 55.29  & 39.47 & 51.34 & 0.02 & 6.31 & 87\\
24 & 148.3  & 39.38 & 51.25 & 0.02 & 5.94 & 71 \\
25 & 283.3  & 39.35 & 51.21 & 0.02 & 5.60 & 65 \\
26 & 75.69  & 39.31 & 51.18 & 0.02 & 6.28 & 60 \\
27 & 238.4  & 39.26 & 51.13 & 0.01 & 5.70 & 54 \\

\hline
\end{tabular}
\end{center}
\end{table*}

\subsection{Abundance estimates}
Optical emission lines from star forming regions  have long been the primary means of gas-phase chemical diagnosis in galaxies (\citealt{pagel}, \citealt{aller90}, \citealt{Shields} and references therein). Optical emission line spectroscopy preferentially samples the warm ionized phase of the interstellar medium in the immediate vicinity of recent star formation events. Chemical analyses of these star forming regions require measurement of H and He recombination lines along with collisionally excited lines from one or more ionization states of heavy element species.  Oxygen is the most commonly used metallicity indicator in the interstellar medium by virtue of its high relative abundance and strong emission lines in the optical part of the spectrum and also due to the fact that oxygen is an excellent coolant. Electron temperature drops with the increasing abundance of coolants and hence oxygen lines are also good indicator of electron temperature. The most precise method of determining the abundance of galaxies requires the detection of temperature sensitive [OIII] $\lambda$4363 \AA, an emission line which is often not detected in the metal rich galaxies.  Abundance estimate of the star forming knots can be made by statistical methods based on strong lines. The abundance calculation is made with various methods available in the literature for optical emission line spectra (\citealt{zaritsky}(Z94), \citealt{mcgaugh}(M91), \citealt{cl}(CL01), \citealt{kobul}, \citealt{kewley02}). We compared the results of these methods and found them to be agreeing well with each other, the average value is being treated as the oxygen abundance as shown in Table \ref{tab6}. The temperature determination could not be made using the standard line ratio  ([OIII] $\lambda$5007 \AA+ $\lambda$4959 \AA)/[OIII] $\lambda$4363 \AA \ as $\rm{[OIII]} \ \lambda4363 \ \AA$ \ was too weak to observe.

From Table \ref{tab6}, it is clear that the mean [NII]/H$\alpha$ ratio for the spectroscopically observed complexes is $\sim$0.3, and the mean [SII]/H$\alpha$ ratio is $\sim$0.25, except for the complex C18, which shows considerably larger value of [SII]/H$\alpha$ ratio. The [SII]/H$\alpha$ ratio can be used as a discriminant for shock heated versus photoionized gas (\citealt{blair}, \citealt{whitmore}). The regions which are heated due to radiative shocks occuring in relatively dense material (as is seen in supernovae remnants)  tend to have $  \rm{[SII]/H}\alpha > 0.4$, whereas the regions which are photoionized by the light from hot stars (i.e. HII regions) are known to have lower values of   [SII]/H$\alpha$ (typically $<$ 0.2). So,  all the spectroscopically observed complexes, except for C5, C16, C18, appear to be purely photoionized. Complex C18 shows considerable differences as compared to  others:  it has a low value of internal reddening, low value of [NII]/[OII] ratio, but with intermediate range in H$\alpha$ luminosity and colours. This unusual behaviour of complex  C18 might  be explained if this region  has a very high supernova explosion rate. The number of supernovae observed in the last 40 yrs in this galaxy has been 3 which is moderately high. These are SN1961P, SN1996an and SN1998dl (the last two are TypeII events). Shocks due to older supernovae in the C18 regions might have ionized the medium and lowered the internal reddening by sweeping out the interstellar material.  

\begin{figure}
\resizebox{\hsize}{!}{\includegraphics{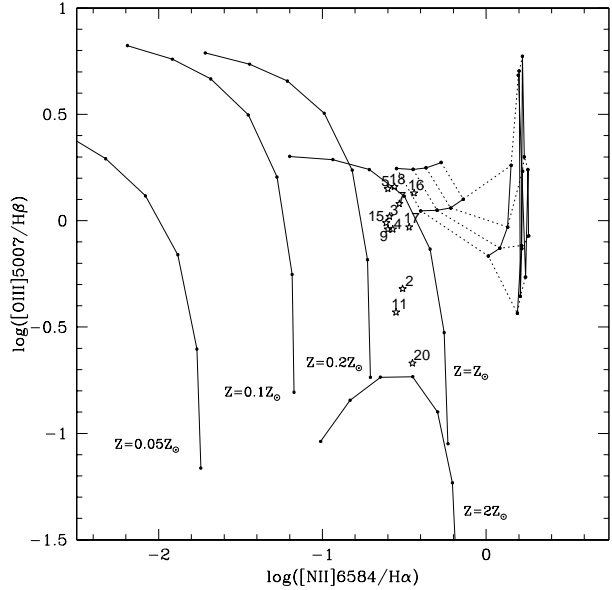}}
\caption[]{Diagnostic diagram of [OIII]/H$\beta$ vs. [NII]/H$\alpha$ for the spectroscopically observed complexes plotted as filled dots. Also plotted are the grids of Mappings III starburst (solid lines) models for different $Z$ and shock (solid and dotted lines to the upper right) models, the dotted lines represent increasing shock velocity $V_s$ = 150, 200, 300, 500, 750 and 1000 kms$^{-1}$ and the connected solid lines are for increasing magnetic parameter $b = 0.5, 1.0, 2.0$ and $4.0$. Plot is similar to that of \cite{mazzuca06}. For explanation on models refer \cite{dopita2000} and \cite{allen98}. }
\label{fig10}
\end{figure}

To further analyse this issue we use the standard diagnostic diagrams of \cite{vo87} modified by \cite{dopita2000} and \cite{allen98} as Mappings III. Figure \ref{fig10} shows the plot of log([OIII]/H$\beta$) versus log([NII]/H$\alpha$) for spectroscopically observed complexes (refer Table \ref{tab6}) along with the Mappings III model prediction for HII regions for an instantaneous burst model adopted from Starburst99. Also plotted in the figure are the model predictions due to shock ionization. Figure \ref{fig10} is similar to Figure 3 of \cite{mazzuca06}.

It is clear from Figure \ref{fig10} that, for the star forming complexes, [OIII]/H$\beta$ and [NII]/H$\alpha$ ratios lie in the solar metallicity region. Although, [NII]/H$\alpha$ values which are the indicators of strong shock excitation or an AGN \citep{dopita02}, vary within a small range of 0.25 to 0.35, the [OIII]/H$\beta$ ratio found to vary in a large range from 0.35 to 1.45 for these complexes. Low values of [OIII]/H$\beta$ characterize the star forming regions and photoionization due to massive OB type stars whereas high values of [OIII]/H$\beta$ are due to other mechanisms like shock ionization. $ \rm{[OIII]/H}\beta > 3$ corresponds to photoionization of gas by non-thermal or power law type continuum radiation \citep{allard06}.  High [OIII]/H$\beta$ values can arise in HII regions if the metallicity of the gas is sufficiently changing within the galaxy but as can be seen here metallicity remains constant over the face of the galaxy and hence the large fluctuations of the [OIII]/H$\beta$ ratio within a galaxy suggest a variation of the ionizing mechanism \citep{sarzi06}.

From Figure \ref{fig10}, complexes C5, C7, C16 and C18 are all having  [OIII]/H$\beta$ $> 1$ and this value is similar to shock ionization in LINER nuclei of galaxies (\citealt{vo87}, \citealt{allard06}). Complex C16 falls right in the shock region of Mappings III model; closely following this complex are C18 and C5. By considering the above facts we can substantiate that complexes C16 and C18 form the shock affected region. From the mixed population colour-colour diagram (refer Figure \ref{fig8}) for complex C16, it is seen that the underlying older population present, is in the age range 10-20 Myr. Also the fraction of young to old component for C16 is 0.1 meaning that young recent burst component is 10\% when compared to the older component or in other words the quantity of older population present is higher. Considering these two facts, one can see that the massive stars of the older population present in this region have undergone supernova explosion whose shells might have expanded sufficiently into space creating shocks. Such shocks can also induce further burst of star formation.

\begin{table*}
\begin{minipage}{146mm}
\caption{Oxygen Abundances of the complexes for which spectra could be obtained.}
\label{tab6}
\centering
\begin{tabular}{l|l|cccccccccccc}
\hline 
   
  & & & & & & & & & & & & \\ \\
 ID & 2 & 3 & 4 & 5 & 7 & 9 & 11 & 15 & 16 & 17 & 18 & 20 \\ \\ 
\hline
\hline \\

 $ E(B-V)$ & 0.34 & 0.29 & 0.45 & 0.10 & 0.32 & 0.44 & 0.50 & 0.17 & 0.14 & 0.53 & 0.07 & 0.47 \\ \\
 $\rm log([NII]/H_{\alpha})$ & -0.51 & -0.59 & -0.57 & -0.60 & -0.53 & -0.60 & -0.55 & -0.61 & -0.44 & -0.47 & -0.56 & -0.45 \\ \\ 
 $\rm log([OIII]/H_{\beta})$ &  -0.32 & 0.02 & -0.04 & 0.15 & 0.08 & -0.04 & -0.43 & -0.01 & 0.13 & -0.03 & 0.16 & -0.67 \\ \\
${\rm log([SII]/H_\alpha})$ & -0.66 & -0.64 & -0.55 & -0.36 & -0.46 & -0.60 & -0.64 & -0.53 & -0.32 & -0.42 & -0.30 & -0.61 \\ \\
${\rm [SII](6717/6731)}$ & 1.48 & 1.38 & 1.48 & 1.53 & 1.33 & 1.35 & 1.85 & 1.34 & 1.43 & 1.48 & 1.47 & 1.13 \\ \\
${\rm log([NII]/[OII])}$ & -0.53 & -0.52 & -0.65 & -0.70 & -0.60 & -0.47 & -0.50 & -0.33 & -0.61 & -0.65 & -0.86 & -0.26 \\ \\
$ log(R_{23})$ & 0.52 & 0.53 & 0.62 & 0.68 & 0.72 & 0.46 & 0.43 & 0.39 & 0.76 & 0.69 & 0.85 & 0.22\\ \\
\hline \\
& & & &  & {\bf log(O/H)+12} & &  & & & & & & \\ \\
\hline \\
$ log{\rm ([NII]/[OII])}$ & 8.89 & 8.898 & 8.82 & 8.67 & 8.85 & 8.92 & 8.91 & 8.98 & 8.85 & 8.83 & 8.69 & 9.02 \\ \\
$ Z94$ & 8.985 & 8.97 & 8.88 & 8.67 & 8.91 & 9.03 & 9.06 & 9.08 & 8.70 & 8.804 & 8.54 & 9.18\\ \\
$M91$ & 8.83 & 8.82 & 8.73 & 8.56 & 8.63 & 8.92 & 8.88 & 8.92 & 8.59 & 8.66 & 8.60 & 8.99\\ \\
$CL01$ & 9.10 & 8.90 & 8.86 & 8.69 & 8.81 & 8.83 & 9.01 & 8.72 & 8.74 & 8.85 & 8.58  & 9.09\\ \\
${\bf Avg}_{Z,M,CL}$ & 8.97 & 8.897 & 8.82 & 8.64 & 8.78 & 8.93 & 8.98 & 8.91 & 8.67 & 8.77 & 8.57 & 9.09\\ \\
\hline
\end{tabular}
\end{minipage}
\end{table*}

\subsection{Star formation rate}
The star formation indicators spanning from ultraviolet to radio are used to obtain the star formation rates in galaxies. The nebular lines effectively re-emit the stellar luminosity shortward of the Lyman limit, so they provide a direct probe of the young massive stellar population. Hydrogen recombination lines, specially H$\alpha$ line flux have been used extensively to estimate the number of Lyman continuum photons that are being emitted by massive stars in the HII regions and star formation rates (\citealt{kenni83}, \citealt{kenni94}, \citealt{kenni98}, \citealt{calzetti05}). Since the main source contributing to  the ionizing radiation are the stars with masses greater than 10 $M_{\odot}$ and age less than 10 Myr, it is expected that the H$\alpha$ flux is proportional to the number of photons produced by the young, hot stars and so the H$\alpha$ flux provides a nearly instantaneous measure of the SFR, independent of the previous star formation history \citep{kenni98}. There are many calibrations available for  conversion of H$\alpha$ luminosity  to SFR (\citealt{kenni83}, \citealt{kenni94}, \citealt{dopita94}, \citealt{leith95}). The most recent calibration by \cite{kenni98} for solar abundances and Salpeter initial mass function (IMF) with the mass range $0.1-100 \ M_{\odot}$ is 
\begin{center}
\begin{equation}
$$\rm{SFR} \ (M_{\odot} \ \rm{{yr}^{-1}}) = 7.9\times 10^{-42} \ L\rm{(H\alpha)\  erg\  s^{-1}}$$
\end{equation}
\end{center}
The observed H$\alpha$  luminosities of the H II regions corrected  for extinction and contribution from the [NII] line, range from 10$^{38.74}$ erg s$^{-1}$ for complex C14 to 10$^{40.82}$ erg s$^{-1}$ for complex C9. Using the calibration by \cite{kenni98} we calculated the SFR for the individual complexes as reported in Table \ref{tab5}. 

Complexes C2, C3, C4, C7, C9 and C11 all have high star formation rate and from Figure \ref{fig3} we can see that these complexes lie in the northern  part of the galaxy in the spiral arms whereas other complexes namely C12-19 which lie in the southern part of the galaxy show a factor of 10 lower star formation. Complexes C20-23 which lie in the nuclear proximity, also show lower star formation. Complexes C12-17 which lie in the southern part of the galaxy, show relatively high H$\alpha$ equivalent width, younger age but low SFR. Also, the CO map reveals high density of the molecular gas \citep{young} in the southern part of the galaxy.  From the observed high H$\alpha$ equivalent width, younger age and very low SFR, it appears that  the south-west part of the galaxy represents young compact star forming regions.  Similar compact star forming regions with extreme equivalent width of H$\alpha$ are found in galaxy clusters  Abell 539 and Abell 634 \citep{rever}. It is proposed that the  compact starburst  in Abell 539 resulted because of the compression of the intergalactic clouds.  We expect that  in the southern part of NGC 1084, where the density of the molecular gas is high \citep{young}, similar compression of the gas could have caused the compact bursts of star formation. 

We have also estimated the integrated H$\alpha$ flux and H$\alpha$ luminosity of the galaxy as $8.32 \times10^{-12}$ erg\  cm$^{-2}$\  s$^{-1}$ and $3.5 \times 10^{41}$ erg \ s$^{-1}$, respectively,  which matches well with their respective estimates  of $6.6 \times10^{-12}$ erg\  cm$^{-2}$\  s$^{-1}$ and $2.78 \times 10^{41}$ erg\ s$^{-1}$ by Kennicutt (1983). The total star formation rate of the galaxy using Eq. $(2)$ turns out to be $2.77 \ M_{\odot}${yr}$^{-1}$. The IRAS flux at 60 $\micron$ is  S$_{60 \micron}=29.7$ Jy \citep{young89}. The corresponding luminosity at 60 $\micron$ using the relation given by  \cite{dever97} is $14.74 \times10^{9}$ $L_\odot$. For galaxies having S$_{100\micron}$/S$_{60\micron}$ colour typical of spiral galaxies, the 60 $\micron$ luminosity represents $\sim$40\% of the total luminosity in the range 1-1000 $\micron$. The infrared luminosity in the range 1-1000 $\micron$ can be used to estimate  star formation rate of the galaxy using relation given by \cite{sco83}. The star formation rate for NGC 1084 using FIR luminosity is 2.84 $M_{\odot}${yr}$^{-1}$, which is in good agreement with  our estimate of SFR using  H$\alpha$ flux. Clearly, the star formation rate is high but not as high so as to classify this galaxy as a starburst.

 NGC 1084 is known to have kinematically distinct regions in the central part (R $< 5''$) and in the periphery of the galaxy at R = $40'' - 50''$ to the north-east from the centre. The central peculiarity was thought to arise due to the presence of a minibar, but in the absence of observational signature of a bar in the optical and $H$ band image it is interpreted as a result of triaxial potential \citep{fridman}. The fast moving region with  non-circular motion in the periphery of the galaxy termed as ``spur'' (shown in Figure \ref{fig3}) extends up to  1-2  kpc and is observed in the regions located in between the HII complexes C3, C4 and beyond, which is accompanied by  sharp rise in the [NII]/H$\alpha$ ratio compared to those of neighbouring HII regions \citep{moiseev}, indicative of the presence of strong shock waves. The higher value of [OIII]/H$\beta$ and also moderate value of [NII]/H$\alpha$ ratio for complex C5 (Figure \ref{fig10}) indicates the influence of a mild shock. The enhanced [NII]/H$\alpha$ ratio and the non-circular motion in the outer region of the galaxy is explained by \cite{moiseev} as due to a possible  infall of extraplanar high-velocity clouds or tidal disruption and accretion  of a small galaxy.  From our analysis, discussed in Section \ref{dis}, southern and eastern regions of the galaxy comprising complexes C12, C13, C14, C15, C16, C18, C19 and C24 are less obscured from dust as is evident from the reddening value which is 0.07 $<$ E(B-V) $<$ 0.15 mag for this region. Hence the south-eastern region (C18, C19, C24, C15, C14) of the galaxy disk is the nearest to the observer.  These regions have very little star formation evident from their star formation rates (refer Table \ref{tab5}). From the colour-colour mixed population diagram (Figure \ref{fig8}), it is also evident that complexes C12, C16 and C24 are forming young stars ($<6 \ \rm{Myr}$) but only in a small fraction ($\sim$10$\%$) compared to the old population already present. Northern complexes namely C2, C3, C4, C6, C7, C8, C9 all have reddening in the range 0.3 $<$ E(B-V) $<$ 0.5 mag, indicating that they lie on the far side of the disk from the observer. \cite{moiseev} detected velocity anomalies in the ``spur'' region, wherein the redshifted residual velocities for the gas are nearer to complexes C2, C3 and blueshifted residual velocities for the gas flow are detected nearer to the complexes C4, C5 and C25 region, indicating an off-center rotating polar ring showing us the receding and approaching parts. From the mixed population evolutionary tracks (Figure \ref{fig8}), we find evidence for a starburst 10 Myr before the current one in several complexes: C5, C6, C12, C16, C18, C24, C25, C26, C27. An earlier starburst separated by 20-30 Myr may be present in complexes C3, C8 and C15. Still older starbursts may be present in complexes C4, C10, C11 and C18, but are not easy to distinguish due to the recent strong burst. However, C1, C17, C20, C21, C22, C23 provide indications of a weak burst superposed on a 200 Myr old population.

The 1.49 GHz map of NGC 1084 obtained with VLA by Condon (1987) shows strong emission coincident with northern complexes (C2, C3, C4, C7, C9). This map shows another radio source $\sim$$3'.5$ south-west of the centre of NGC 1084, and also a bridge connecting the disk of NGC 1084. A small island of strong H$\alpha$ source \citep{moiseev} lies on this bridge and also a blue region seen in our $B-V$ map (Figure \ref{fig2}). \cite{moiseev} proposed that NGC 1084 may have interacted and merged with a gas rich dwarf galaxy recently. A recurrent burst of star formation may have occurred in the circumnuclear region 200 Myr ago due to this interaction.  Our mixed population models, show recurrent starbursts over the last 40 Myr. Deeper optical, IR and HI observations are needed to ascertain the nature of \cite{condon87} source and its relationship with NGC 1084.

\section{Conclusion}
\label{con}

Broad band $U$$B$$V$$R$$I$, narrow band H$\alpha$ photometric and spectroscopic studies of the star forming regions of NGC 1084 are presented.  The total star formation rate in this galaxy is found to be $2.8$ $M_{\odot}${yr}$^{-1}$, not so high as to classify this galaxy as a starburst galaxy. Star formation in this galaxy is highly chaotic wherein the star formation is not confined to the spiral arms only. This galaxy is forming stars at different rates in the northern and southern regions. Northern part of the galaxy shows bulk production of massive stars as is evident from the SFR in the range of $0.3 <$ SFR $< 0.5$ $M_{\odot}${yr}$^{-1}$ for individual complexes. Southern part of the galaxy shows much lower star formation and at a much lower proportion compared to the earlier bursts. Ages of recently formed young stars in the southern region are $<4$ Myr whereas the northern complexes have ages in the range 5-6 Myr. Bulk production of stars in the northern region is attributed to a series of shorter bursts of star formation over the past 40 Myr, possibly due to interaction and merger with a gas rich dwarf galaxy. There is a radio source $3'.5$ south-west of the galaxy, connected to it by a bridge \citep{condon87}, and it would be of interest to understand the nature of this source.

Differences found in the star formation parameters in the northern and southern part of the galaxy disk are not seen in metallicity. All the spectroscopically observed complexes show metallicities close to solar.

\section*{Acknowledgements}
We would like to thank the anonymous referee for comments and suggestions provided to improve the quality and presentation of the paper. This work has made use of the NASA Astrophysics Data System and the NASA/IPAC Extragalactic Database (NED) which is operated by Jet Propulsion Laboratory, California Institute of Technology, under contract with the National Aeronautics and Space Administration.
RS would like to thank University Grants Commission, GOI for their financial assistance.
Authors wish to thank the support given by the staff of IAO and CREST, IIA.

\end{document}